\documentclass[conference]{IEEEtran}
\IEEEoverridecommandlockouts

\usepackage{cite}
\usepackage{amsmath,amssymb,amsfonts}
\usepackage{algorithmic}
\usepackage{graphicx}
\usepackage{textcomp}
\usepackage{xcolor}

\usepackage{microtype}
\usepackage{caption}
\usepackage{subcaption}
\usepackage{multirow}
\usepackage[hidelinks]{hyperref}
\usepackage{eso-pic}

\makeatletter
\def\@maketitle{%
	\newpage
	\bgroup
	\par\vskip\IEEEtitletopspace\vskip\IEEEtitletopspaceextra
	\centering%
	\vskip0.2em
	{\Huge\@title\par}%
	\vskip1.0em\par%
	\mbox{}\hfill\begin{@IEEEauthorhalign}\@author\end{@IEEEauthorhalign}\hfill\mbox{}\par%
	\egroup
	\par\addvspace{0.4\baselineskip}
}
\makeatother

\makeatletter
\def\ps@IEEEtitlepagestyle{%
	\def\@oddfoot{\mycopyrightnotice}%
	\def\@oddhead{\hbox{}\@IEEEheaderstyle\leftmark\hfil\thepage}\relax
	\def\@evenhead{\@IEEEheaderstyle\thepage\hfil\leftmark\hbox{}}\relax
	\def\@evenfoot{}%
}
\def\mycopyrightnotice{%
	\begin{minipage}{\textwidth}
		\scriptsize
		Copyright~\copyright~2025 IEEE. DOI: 10.1109/ICCAD66269.2025.11240659. Personal use of this material is permitted. 
		Permission from IEEE must be obtained for all other uses,
		in any current or future media, including reprinting/republishing this material
		for advertising or promotional purposes, creating new collective works,
		for resale or redistribution to servers or lists, or reuse of any copyrighted
		component of this work in other works by sending a request to pubs-permissions@ieee.org.
	\end{minipage}
}
\makeatother

\def\BibTeX{{\rm B\kern-.05em{\sc i\kern-.025em b}\kern-.08em
    T\kern-.1667em\lower.7ex\hbox{E}\kern-.125emX}}
\begin{document}

\title{Rhea: a Framework for Fast Design and Validation\\of RTL Cache-Coherent Memory Subsystems

\thanks{The framework is named after Rhea, Saturn’s second-largest moon, itself named for the Titan Rhea, mother of the first generation of Olympian gods in Greek mythology.}}

\author{\IEEEauthorblockN{Davide Zoni, Andrea Galimberti, Adriano Guarisco}
\IEEEauthorblockA{\textit{Dipartimento di Elettronica, Informazione e Bioingegneria (DEIB), Politecnico di Milano},
Milano, Italy \\
Email: davide.zoni@polimi.it, andrea.galimberti@polimi.it, adriano.guarisco@polimi.it}
}

\maketitle

\begin{abstract}
Designing and validating efficient cache-coherent memory subsystems is a critical yet complex task in the development of modern multi-core system-on-chip architectures.
Rhea is a unified framework that streamlines the design and system-level validation of RTL cache-coherent memory subsystems. On the design side, Rhea generates synthesizable, highly configurable RTL supporting various architectural parameters. On the validation side, Rhea integrates Verilator's cycle-accurate RTL simulation with gem5’s full-system simulation, allowing realistic workloads and operating systems to run alongside the actual RTL under test.
We apply Rhea to design MSI-based RTL memory subsystems with one and two levels of private caches and scaling up to sixteen cores. Their evaluation with 22 applications from state-of-the-art benchmark suites shows intermediate performance relative to gem5 Ruby’s MI and MOESI models.
The hybrid gem5-Verilator co-simulation flow incurs a moderate simulation overhead, up to 2.7 times compared to gem5 MI, but achieves higher fidelity by simulating real RTL hardware. This overhead decreases with scale, down to 1.6 times in sixteen-core scenarios. These results demonstrate Rhea’s effectiveness and scalability in enabling fast development of RTL cache-coherent memory subsystem designs.
\end{abstract}


\section{Introduction}
\label{sec:introduction}
An efficient cache-coherent memory subsystem, composed of
a set of caches, an on-chip interconnect, and memory,
is critical to the performance of modern multi-core SoCs~\cite{Sorin_2022Springer},
which are increasingly getting larger and more complex,
e.g., by integrating hardware accelerators~\cite{Galimberti_2023ICECS} as well as
security-~\cite{Zoni_2025TC,Galimberti_2024ICECS} and power-related components~\cite{Zoni_2023CSUR}.
When designing a new cache-coherent memory subsystem, it is crucial to evaluate both
the functional correctness and the performance of the design without having to implement the entire system,
including both the hardware and software layers, especially when
assessing feasibility requirements or making early-stage design decisions.

The ability to conduct performance evaluations using real-world applications,
such as multi-threaded workloads and scenarios involving multiple concurrent processes, is essential.
By doing so from the early stages, designers can gain insights into
how the cache-coherent memory subsystem will handle complex, realistic workloads while
keeping costs low and avoiding the overhead associated with full-system implementation.

Compact models for cache, interconnect, and memory have been developed and embedded within
system-level simulators such as gem5~\cite{Binkert_2011SIGARCH} and
Sniper~\cite{Carlson_2011SC} to address these challenges.
However, these models lack the accuracy from both the performance and power consumption
standpoints to effectively drive the early design stages.

To overcome the limitations of traditional simulators, the research community has increasingly
turned to prototype-based methodologies that leverage FPGAs~\cite{Balkind_2020Micro,Amid_2020Micro}.
These platforms offer higher fidelity and performance for architectural evaluation,
making them attractive for studying components such as cache-coherent interconnects.
Despite their advantages, developing cache coherence subsystems on FPGA-based prototypes is a challenging task.
Existing solutions are typically ad-hoc and highly specialized,
making them difficult to extend or adapt for evolving architectural requirements~\cite{Tornero_2023DCIS}.
Moreover, the need to run complete operating systems (OSes) and real-world applications to
effectively verify the correctness and evaluate the quality of cache-coherent interconnects
drastically limits the usability of prototype-based approaches.
The development process is further complicated by the significant expertise required and
the need for expensive hardware.

Hybrid simulation techniques have emerged as a promising alternative~\cite{Lopez-Paradis_2021ICPP,Fu_2024FPL},
integrating register-transfer level~(RTL) components into system-level simulators like gem5.
These approaches simulate the majority of the system at a high level of abstraction while offering a detailed simulation for specific components either by leveraging RTL simulators or FPGA emulation.
However, the existing approaches are custom-tailored to accelerator designs,
which often function independently from the rest of the system.
Unlike accelerators, cache coherence subsystems are deeply intertwined with the memory hierarchy,
involving frequent and complex interactions across the entire system.
This complexity makes RTL modeling and verification significantly more challenging,
underscoring the need for innovative solutions that balance simulation fidelity with practical feasibility.

There is a growing need for a unified framework that tightly integrates the design and validation of RTL cache-coherent memory subsystems. Such subsystems must be efficient and highly configurable, supporting a wide range of architectural parameters, including the number of cache levels, the hierarchy of cache controllers, and the number of cores. Validation of the designed cache-coherent memory subsystem must encompass both functional verification and detailed evaluation of power, energy, performance, and timing characteristics under the execution of realistic workloads, including full OSes and arbitrary applications, and at cycle-level accuracy.

\begin{figure*}
	\centering
	\includegraphics[width=\textwidth]{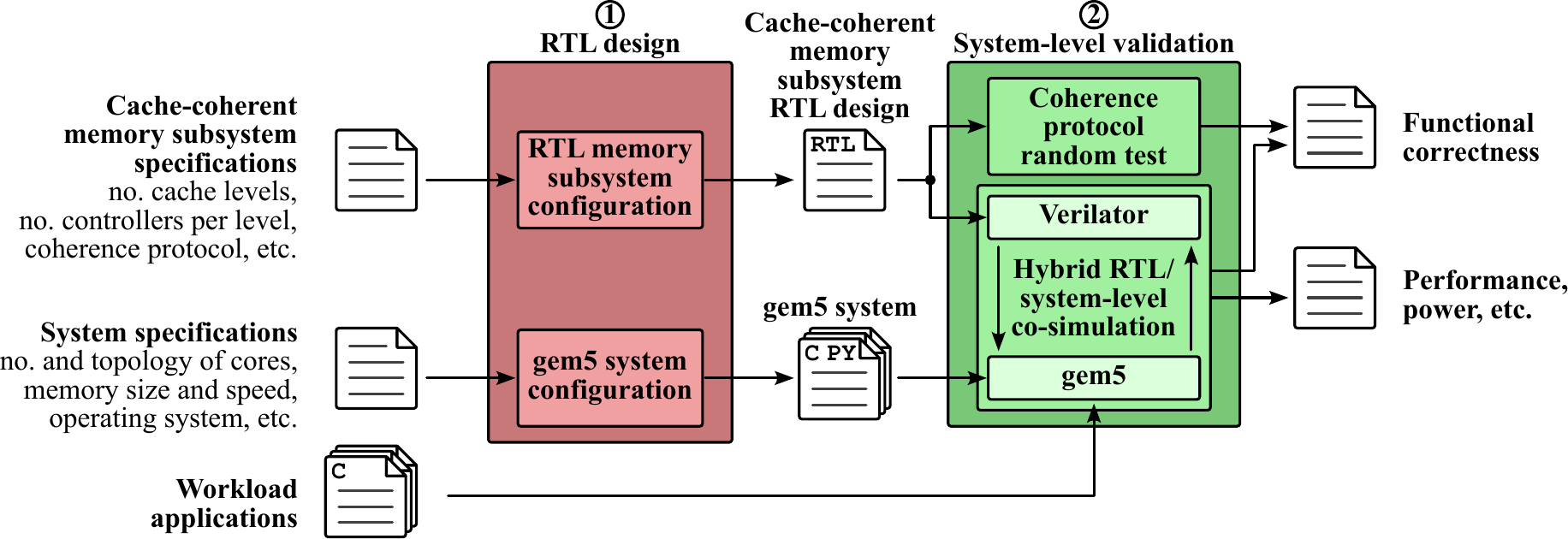}
	\vspace{0.1cm}
	\caption{High-level flow of Rhea framework.}
	\label{fig:flow}
\end{figure*}

\subsection*{Contributions}
\label{ssec:intro_contributions}
To address these challenges, we present Rhea, a unified framework that streamlines the design and validation of RTL cache-coherent memory subsystems. On the design side, it enables the generation of synthesizable RTL that is both efficient and highly configurable, supporting a wide range of architectural parameters, including different numbers of cores, multiple cache levels, and various cache coherence protocols. On the validation side, the Rhea framework integrates RTL and system-level simulation through the widely adopted tools gem5~\cite{Binkert_2011SIGARCH} and Verilator~\cite{Snyder_2018ORConf}, allowing realistic workloads to be executed within a full-system context.

By enabling co-simulation of complex software stacks, including full OSes and multi-threaded applications, alongside detailed RTL components, our framework allows designers to validate functional correctness and analyze power, performance, and timing, even when other parts of the system are still under development. This tight integration of configurable RTL design and system-level validation brings early visibility into both architectural and microarchitectural behavior within a realistic execution environment.

Rhea advances RTL cache-coherent memory subsystem design and validation through three key contributions.

\begin{enumerate}
\item\emph{Configurable RTL design flow} Rhea provides a modular RTL design methodology that supports a broad range of memory subsystem configurations. The generated RTL is fully synthesizable and parameterized, allowing designers to instantiate different numbers of cores, single- or multi-level cache hierarchies, and multiple coherence protocols.

\item\emph{Hybrid RTL/system-level co-simulation} Rhea integrates Verilator-generated models of RTL components into the gem5 system simulator, enabling co-simulation of realistic, full-system software workloads with cycle-accurate RTL hardware models. This approach supports both functional verification and detailed performance, power, and timing analysis.

\item\emph{SystemVerilog coherence protocol tester} Rhea includes a SystemVerilog port of gem5’s Ruby random tester for standalone RTL verification. This tool stress-tests cache-coherent memory subsystems using randomized, high-concurrency memory access patterns to complement application-driven validation.
\end{enumerate}

We demonstrate the effectiveness of the Rhea framework through an
\emph{extensive experimental campaign} that includes single-level and two-level
cache-coherent memory subsystems implementing the MSI protocol and targeting
support for up to sixteen cores. The experimental results show the performance
of the RTL cache-coherent memory subsystem obtained by applying the proposed
framework, along with an analysis of the simulation overhead introduced by
the hybrid co-simulation.

Rhea is publicly \emph{released as open source} in order to foster further research and reproducibility. Rhea's
open-source release includes the designs of SystemVerilog cache-coherent memory subsystems,
the hybrid co-simulation flow, and the SystemVerilog coherence protocol tester.

\section{Related Work}
\label{sec:related}
Full-system simulators~\cite{Binkert_2011SIGARCH,Carlson_2011SC,Ubal_2012PACT}
are flexible and easy to use, however their high level of abstraction hinders 
their accuracy in estimating performance and more prominently power consumption,
due to the need to rely for the latter on event-count-based frameworks, e.g.,
CACTI~\cite{Wilton_1996JSSC} and McPAT~\cite{Li_2013TACO}.
More recent proposals~\cite{Shao_2016MICRO,Iordanou_2019FCCM,Spencer_2024JSA}
specifically target heterogenous SoCs with hardware accelerators.

RTL simulators such as Cadence Xcelium, Siemens ModelSim, and Synopsys VCS
provide the best estimation in terms of performance of the simulated system,
however operating at signal-level granularity drastically lengthens their simulations.
The open-source Verilator~\cite{Snyder_2018ORConf} does not carry out a traditional event-driven RTL simulation
like the aforementioned solutions, but it performs instead a cycle-accurate simulation of
the C++ model compiled from the RTL description, significantly speeding up
the simulation at the cost of a lower accuracy.

A number of frameworks accelerate instead cycle-accurate full-system simulations of
multi-core processors by using FPGAs~\cite{Chiou_2007MICRO,Tan_2010DAC,Pellauer_2011HPCA,Khan_2012ISPASS}.
FireSim~\cite{Karandikar_2018ISCA} notably leverages cloud-hosted FPGAs and
makes use of Chisel to describe the hardware designs,
while commercial solutions such as Cadence's Palladium Z3 can emulate ASIC designs with tens of billion of gates.
Verification approaches based on FPGA emulation in general require
very large FPGA setups, hindering their adoption due to their remarkably high financial cost.

Frameworks such as ESP~\cite{Mantovani_2020ICCAD}, Chipyard~\cite{Amid_2020Micro},
OpenPiton~\cite{Balkind_2020Micro}, BlackParrot~\cite{Petrisko_2020Micro}, and Vespa~\cite{Montanaro_2024ICCD,Montanaro_2025ISCAS}
have emerged for the agile development of heterogeneous multi-core SoCs, and
recent proposals~\cite{Tornero_2023DCIS,Sajin_2023VLSID,
Jung_2024ISCA} specifically target cache coherence features.
All these SoC frameworks, although configurable in some aspects,
deliver rigid architectures that cannot be easily changed,
constrain system designers at an early design stage to
specific OS, compiler toolchain, and hardware target,
and require adopting large FPGAs and expensive boards.

Modern hybrid solutions combine full-system simulations of large SoCs with
higher-accuracy RTL simulations of specific parts of the overall system~\cite{Feng_2017CAL}.
PAAS~\cite{Liang_2017FPL} pairs gem5 and Verilator, that run concurrently and
independently and communicate via inter-process communication, whose large overhead makes it
only suitable to simulate FPGA-based accelerators that
rarely exchange data with the rest of the system and run in parallel with the former most of the time.
gem5+RTL~\cite{Lopez-Paradis_2021ICPP} integrates gem5 and Verilator more tightly to
enable communication between the RTL design under verification and
any of the SoC components via standard gem5 timing ports and packets.
Chimera \cite{Fu_2024FPL} combines gem5 with the possibility to deploy the RTL designs under verification
on FPGA and have them interact with the former via PCIe.
Each RTL IP has indeed a virtual model within gem5 to interact with the simulated SoC,
enabling a transparent forwarding of packets between gem5 and the FPGA.
This approach is only well suited to verifying RTL designs that rarely interact with
the rest of the system, running them in parallel with gem5 and minimizing the communication overhead.

\begin{figure}
	\centering
	\includegraphics[width=\columnwidth]{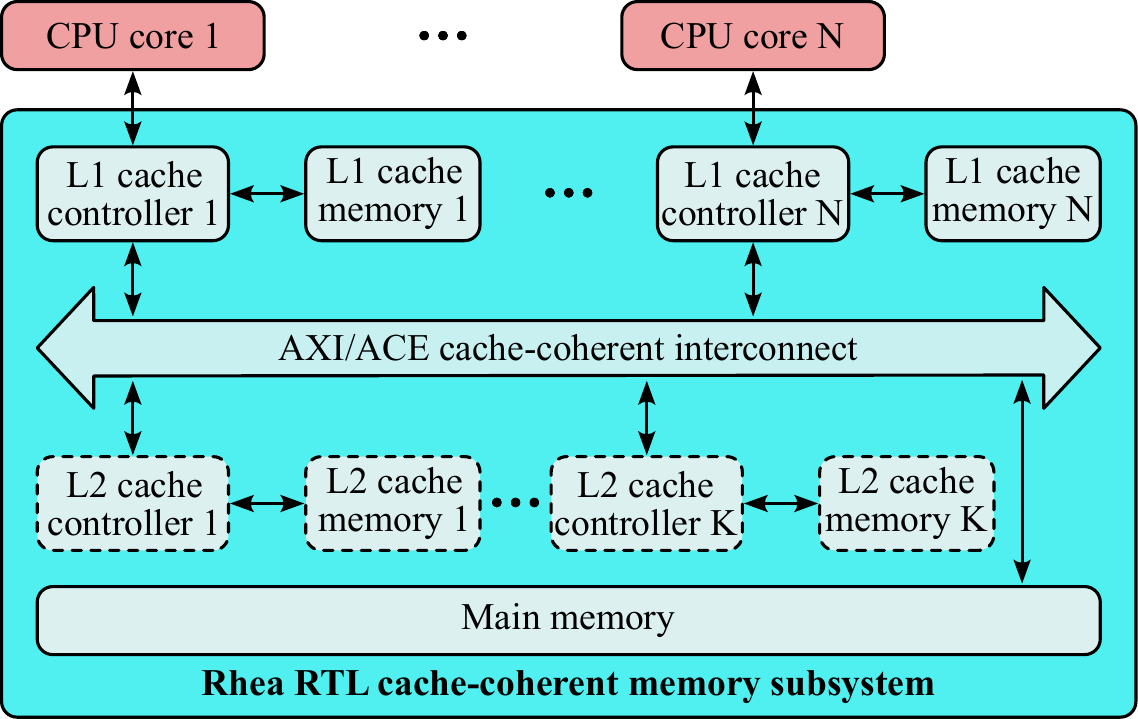}
	\caption{Generic RTL cache-coherent memory subsystem generated by Rhea.
		The number N of L1 caches and the number K of L2 caches are parameters configured at design time.
		L2 caches, denoted by dashed lines, are optionally enabled.}
	\label{fig:design}
\end{figure}

\section{Rhea Framework}
\label{sec:methodology}
The Rhea framework, depicted in Figure \ref{fig:flow}, speeds up developing RTL cache-coherent memory subsystems by
integrating a configurable design process with system-level validation.

The design component of Rhea enables obtaining RTL cache-coherent memory subsystems that are synthesizable and that are configured at design-time by the framework's user to support different numbers of cores, single- or multi-level cache hierarchies, and multiple coherence protocols.

The system-level validation combines a hybrid RTL/system-level co-simulation approach with standalone coherence protocol random testing to verify the functional correctness and validate the performance, power consumption, and other quality metrics of the RTL design. The hybrid co-simulation notably allows stressing the RTL design with real application workloads being executed within the context of the overall system, with also the ability to boot an OS.

The fast design and validation capabilities provided by the Rhea framework enable quickly iterating until the obtained RTL cache-coherent memory subsystems satisfy the designers' requirements and constraints.

\begin{figure*}
	\hfill
	\begin{subfigure}[b]{0.49\textwidth}
		\centering
		\includegraphics[width=0.85\textwidth]{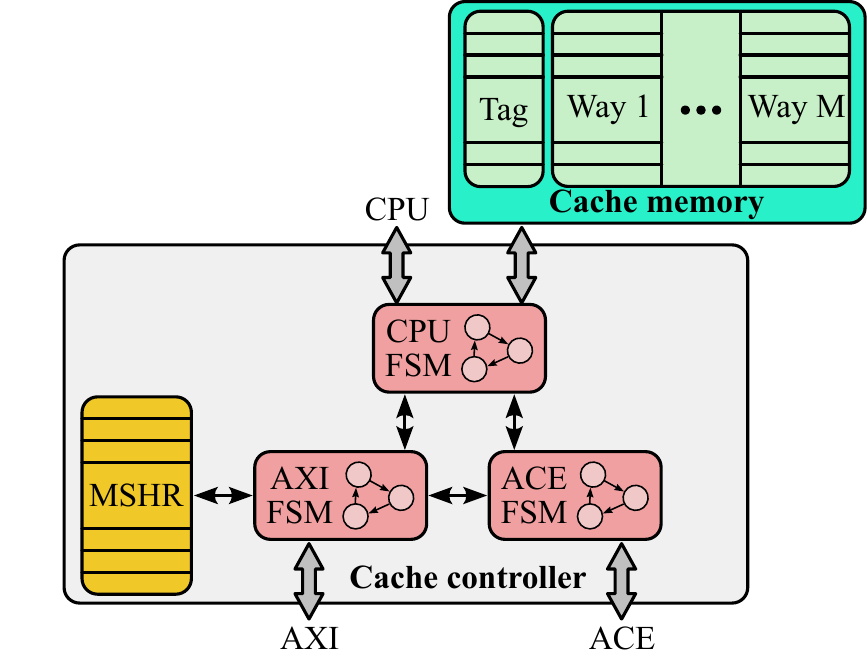}
		\caption{L1 cache controller}
		\label{sfig:template_arch_cache}
	\end{subfigure}
	\hfill
	\begin{subfigure}[b]{0.49\textwidth}
		\centering
		\includegraphics[width=0.85\textwidth]{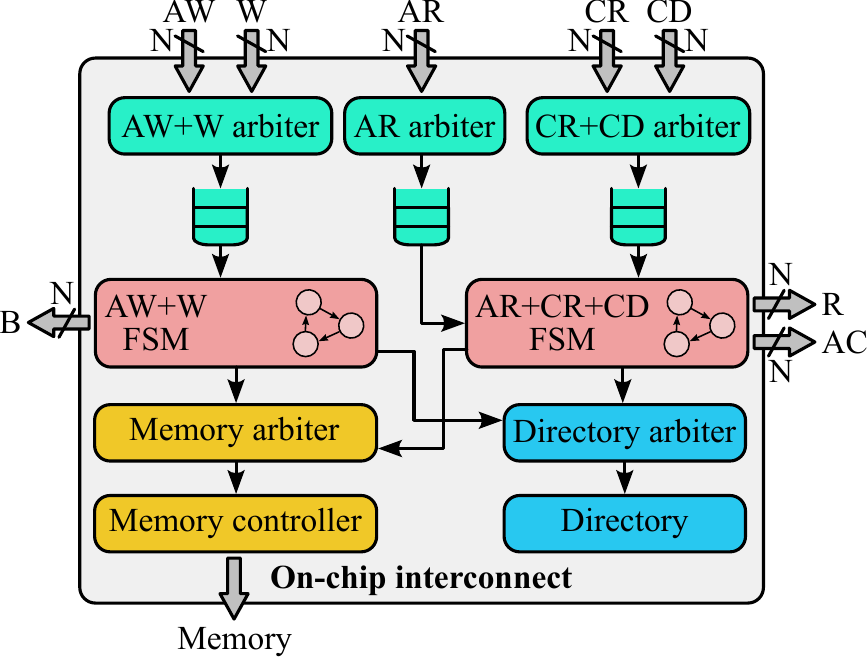}
		\caption{On-chip interconnect}
		\label{sfig:template_arch_iconn}
	\end{subfigure}
	\hfill
	\caption{Detailed architecture of the RTL cache-coherent memory subsystem.}
	\label{fig:template_arch}
\end{figure*}

\subsection{Cache-Coherent Memory Subsystem Design}
The framework enables the designer to configurably generate RTL cache-coherent memory subsystems,
described in SystemVerilog, that support different cache coherence protocols and
up to two levels of caches.
Each generated cache-coherent memory subsystem, as shown in Figure \ref{fig:design},
includes L1 caches and L1 cache controllers, an AXI/ACE bus interconnect,
a memory, and optionally L2 caches and L2 cache controllers.

The L1 cache controller, depicted in \figurename~\ref{sfig:template_arch_cache},
is highly configurable at design time to implement different cache sizes and associativities.
Its CPU interface handles receiving load and store requests and returning their results,
the AXI interface allows the exchange of coherent requests and responses with the on-chip interconnect, and
the ACE interface enables interacting with the on-chip interconnect when snooping activity is in progress.
The L1 controller includes a miss status holding register (MSHR),
that holds the pending request, and three finite state machines (FSMs).
An FSM handles the CPU requests and issues coherent requests,
while the other two manage the AXI coherent transactions and
the ACE snoop transactions, respectively.

The coherent interconnect, whose architecture is shown in \figurename~\ref{sfig:template_arch_iconn},
connects L1 controllers and main memory and is highly configurable via
design-time parameters to adapt to different system requirements,
e.g., core count and AXI data bus width.
Three AXI/ACE channel round-robin arbiters serialize requests issued by L1 cache controllers.
Five first-in, first-out (FIFO) queues, one per AXI/ACE channel and parametrically sized to
avoid back-pressure build-up, store requests until interconnect is ready to serve them.
The directory tracks cache lines present in the system and the state and sharers of each cache line.
The memory controller interfaces the on-chip interconnect with the main memory by serializing and deserializing cache line data.
A dedicated FSM manages the evict and write-back requests issued by the L1 cache controllers
on the AW and W channels of the AXI protocol which require no snooping action,
while another FSM handles the read-clean and read-unique requests and their snooping activity sent via
the AR AXI channel, which often require the interconnect to snoop one or more of the other cache line sharers.
The memory and directory arbiters manage concurrent accesses to the memory and the directory by the two FSMs.

The framework can optionally produce a two-level cache-coherent memory subsystems by also instantiating L2 caches and their controllers. The L2 cache controllers and their associated memory structures follow a similar architectural design to those used for L1 caches and depicted in Figure \ref{sfig:template_arch_cache}. However, in contrast to the private per-core L1 caches, each L2 cache is shared among multiple cores. L2 cache controllers are connected exclusively to the cache-coherent interconnect, serving requests from multiple L1 caches via coherence transactions. The number of L2 cache memories and controllers, and conversely the number of cores sharing each L2, as well as cache parameters such as associativity and capacity, are all configurable design-time options within the proposed Rhea framework.

The cache coherence protocol supported by the memory subsystem is notably configured by applying
simple modifications of the FSMs in the interconnect and cache controller components,
as well as eventually extending both the MSHR with additional transient states and
the directory entries with the corresponding additional bits. 

\subsection{System-Level Validation}
The system-level validation of the designed RTL cache-coherent memory subsystem,
as depicted in \figurename~\ref{fig:flow}, spans the full-system and RTL domains
by including both a hybrid RTL/system-level co-simulation and an RTL cache coherence
protocol random test.

On the one hand, the RTL design is compiled into a C++ model for
Verilator's cycle-accurate simulation that is then integrated into gem5 as
the cache-coherent memory subsystem of a full system which is
able to run applications on top of an OS.
The gem5-Verilator integrated simulation allows verifying
the correctness of arbitrary applications selected by the user and
evaluating the performance of the overall cache-coerent system when executing those workloads.
In addition, the RTL/system-level simulation checks the correctness of
each single operation involving memory that are executed by the various cores
by comparing the responses received by the RTL cache-coherent memory subsystem with
those from the still included Ruby memory model. 

On the other hand, a traditional RTL simulation of the cache-coherent memory subsystem
stress-tests the latter by connecting it to a SystemVerilog port of gem5's Ruby random tester,
enabling the verification of its behavior under randomized and
potentially unpredictable memory access scenarios.

\begin{figure}
	\centering
	\includegraphics[width=\columnwidth]{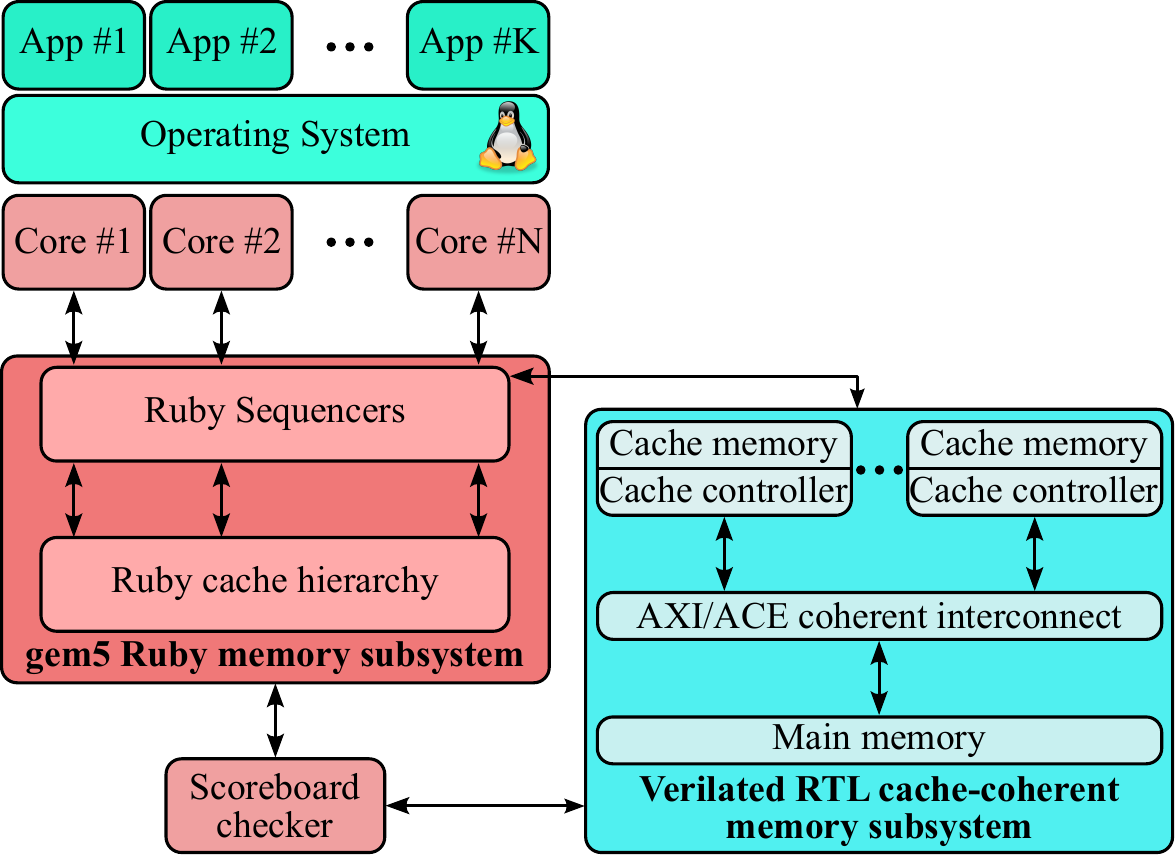}
	\caption{Detailed overview of the integrated gem5-Verilator RTL/system-level simulator.}
	\label{fig:infrastructure}
\end{figure}

\subsubsection{Hybrid RTL/System-Level Co-Simulation}
At startup, the gem5 simulation instantiates the Verilator-generated (or verilated) model of
the RTL cache-coherent memory subsystem under verification and
initializes the input and clock signals and performs a reset sequence by asserting the reset signal,
calling the \texttt{eval()} function to update the model, and then
deasserting the reset signal and invoking \texttt{eval()} again.
The integration of the verilated RTL model into gem5, as shown in Figure \ref{fig:infrastructure},
also utilizes a scoreboard for result comparison and per-CPU queues to manage memory requests.
Ruby Sequencer objects intercept all the CPU-to-cache and cache-to-CPU packets.
CPU-to-cache packets are queued for the RTL model while also being forwarded to Ruby,
with a corresponding entry created in the scoreboard, while
cache-to-CPU packets are held until the verilated model's results are available,
allowing for result comparison before forwarding the packet to the CPU.

At the beginning of each gem5 clock cycle, the outputs of the Verilator model are scanned to detect
responses to previous requests.
If an acknowledgment from the DUT is detected, i.e., its \texttt{ack} signal is equal to 1,
the output signals of the verilated model are processed and the result is saved to the scoreboard.
The corresponding input signals are then reset and the fulfilled request is popped from the queue.
Then, if there are some requests in the queues and the corresponding L1 cache controller is not busy,
a new request is sent to the RTL model.
At the end of the gem5 clock cycle, the verilated model is stimulated by toggling the clock
signal from 0 to 1 and then from 1 to 0 and calling the \texttt{eval()} function to propagate the signal changes.
If there are no pending requests in the system this step is instead skipped to speed up the simulation.
Such loop continues until simulation has completed.

Verilator's ability to compile the RTL design into multithreaded C++ can be leveraged to further reduce the portion of simulation time required by the synchronization and data exchange between gem5 and the verilated model and by the simulation of the verilated model itself.
In particular, let M be the number of threads of the multithreaded verilated model, one thread is used to read the inputs and set the outputs at the boundaries with gem5 while the remaining M-1 threads are used to parallelize the simulation of the DUT.

\subsubsection{RTL Cache Coherence Protocol Random Test}
In addition to the hybrid co-simulation, a complementary verification of
the RTL design is carried out in a traditional RTL simulator by means of
a SystemVerilog port of gem5's Ruby random tester.

The latter is a verification tool included in gem5 to stress-test and verify,
under randomized and potentially unpredictable memory access scenarios,
the correctness of the cache coherence protocol implementation of
memory systems modeled by gem5’s Ruby component, which provides
a highly detailed and flexible framework for simulating
various memory hierarchies and coherence protocols.
The SystemVerilog port of the Ruby random tester included in the proposed framework can
be used directly on the RTL implementation to verify the correctness of
the latter with the same access patterns used in the original gem5 C++ version
of the Ruby random tester.

In particular, the Ruby random tester picks random checks from
a collection of checks and submits them to a random L1 cache controller.
Each check consists of four write requests to consecutive bytes
followed by a read operation to verify the correctness of the result.
Different L1 cache controllers might act on different bytes of the same check,
stressing the cache coherence of the design under test.

\section{Experimental Evaluation}
\label{sec:experiments}
The experimental evaluation assesses both the RTL cache-coherent memory subsystems generated using Rhea and the RTL/system-level co-simulation flow. Specifically, Rhea is applied to design MSI-based RTL memory subsystems featuring one and two levels of private caches and supporting up to sixteen CPU cores. The execution of 22 applications from state-of-the-art benchmark suites enables a quantitative analysis of the performance delivered by the RTL designs and the simulation overhead introduced by the gem5-Verilator integration, relative to standard gem5 simulations.

\subsection{Use Cases}
\label{ssec:exp_usecase}
As representative use cases of the adoption of the proposed Rhea framework,
we design and validate eight instances of RTL cache-coherent memory subsystems,
whose design is described in SystemVerilog and synthesizable, 
that implement the same MSI cache coherence protocol.
The eight instances combine support for 
two different numbers of levels of caches, namely, of one and two,
and for four different numbers of CPU cores, namely, of two, four, eight, and sixteen.

Both single- and two-level MSI RTL cache-coherent memory subsystems are parametrically configured to
feature 8kB private L1 caches, one per each connected CPU core and with four-way set associativity,
an AXI/ACE interconnect with a data width of 32 bits, and a 1GB main memory.

The two-level MSI instances accordingly include L2 caches and the corresponding L2 cache controllers.
Notably, they feature two shared L2 caches regardless of the number of connected CPU cores.
Each L2 cache is sized at 256kB with eight-way set associativity,
providing ample capacity and helping to reduce conflict misses across multiple cores.

\subsection{Experimental Setup}
\label{ssec:exp_setup}
The proposed RTL/system-level co-simulation flow makes use of
the gem5 v23.1~\cite{Binkert_2011SIGARCH} and Verilator 4.104~\cite{Snyder_2018ORConf} software versions.
It was run on a server that features
a sixteen-core Intel Xeon Gold 6326 CPU with a 24 MB cache operating at a 2.90 GHz clock frequency
and a 128GB, 3200 MT/s, quad-channel DDR4 memory
and that runs the Rocky Linux 8.10 OS with a Linux 4.18.0 kernel.
The RTL simulation with the SystemVerilog random tester was
instead carried out in the xsim RTL simulator included in AMD Vivado 2024.2.

The target simulated system
implements twelve different configurations that combine
four different numbers of CPU cores and three different cache coherence protocols.
Its architecture includes a dual-, quad-, octa-, or sixteen-core CPU with cores based on the x86 ISA.
The cores, that implement gem5’s timing CPU model, run at a 100MHz clock frequency and are
connected with each other through a point-to-point interconnect.
The Ruby cache memory and coherence model is configured in
a two-level cache hierarchy implementing a MESI or MOESI cache coherence protocol
with L1 caches private to each core, L2 caches shared among them, and
a 1GB, 1600 MT/s, single-channel DDR3 memory.
The Ruby memory hierarchy can also be configured to implement a single-level MI cache coherence protocol.
The simulated system runs the Ubuntu 18.04 OS with a Linux 5.4.49 kernel.

\begin{table}[t]
	\centering
	\caption{Multi-threaded applications from the PARSEC~\cite{Bienia_2008PACT} and Splash-3~\cite{Sakalis_2016ISPASS}
		benchmark suites executed on top of the Ubuntu 18.04 OS on the simulated systems.
		Legend: \textbf{ID} abbreviation used in Figures \ref{fig:perf_protocol} and \ref{fig:perf_flow}.}
	\begin{tabular}{|ll|ll|ll|}
		\hline
		\multicolumn{2}{|c|}{\textbf{PARSEC}~\cite{Bienia_2008PACT}} & \multicolumn{4}{|c|}{\textbf{Splash-3}~\cite{Sakalis_2016ISPASS}} \\
		\hline
		\textbf{Application} & \textbf{ID} & \textbf{Application} & \textbf{ID} & \textbf{Application} & \textbf{ID} \\ \hline
		blackscholes         & BS          & barnes               & BA          & ocean\_ncp           & ON          \\
		canneal              & CA          & cholesky             & CH          & radiosity            & RD          \\
		dedup                & DD          & fft                  & FF          & radix                & RX          \\
		ferret               & FR          & fmm                  & FM          & raytrace             & RT          \\
		fluidanimate         & FA          & lu\_cb               & LC          & volrend              & VR          \\
		freqmine             & FQ          & lu\_ncb              & LN          & water-nsquared       & WN          \\
		streamcluster        & SC          & ocean\_cp            & OC          & water-spatial        & WS          \\
		swaptions            & ST          &                      &             &                      &             \\ \hline
	\end{tabular}
	\label{tab:apps}
\end{table}

The workload executed on the simulated system consists of 22 applications from
the PARSEC~\cite{Bienia_2008PACT} and Splash-3~\cite{Sakalis_2016ISPASS} benchmark suites.
Such applications, listed in Table \ref{tab:apps}, are executed on the simulated system
in a multi-threaded fashion on top of the Ubuntu OS.
In particular, each application is parallelized on a number of threads that is equal to
the number of CPU cores of the target system, i.e., two, four, eight, or sixteen depending on the configuration.

\begin{figure*}
	\begin{minipage}[b]{\textwidth}
		\centering
		\begin{subfigure}[b]{\columnwidth}
			\centering
			\includegraphics[width=\textwidth]{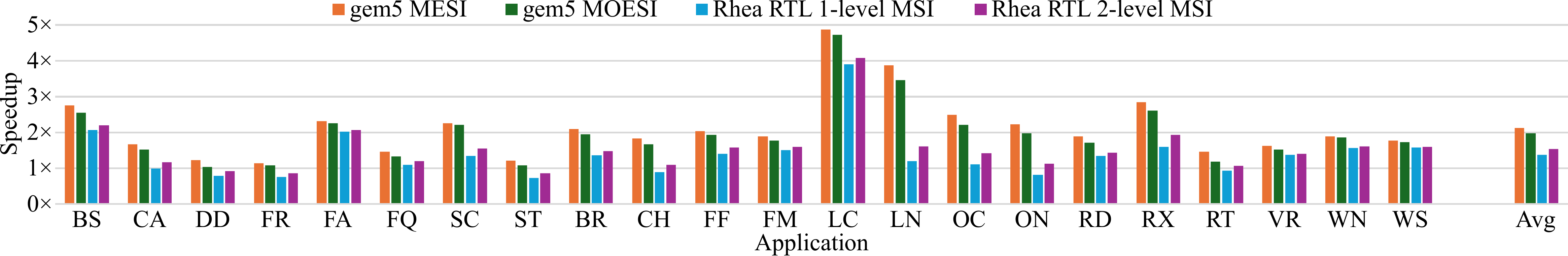}
			\caption{Dual-core configuration}
			\label{sfig:perf_protocol-2}
			\vspace{0.1475cm}
		\end{subfigure}
		\begin{subfigure}[b]{\columnwidth}
			\centering
			\includegraphics[width=\textwidth]{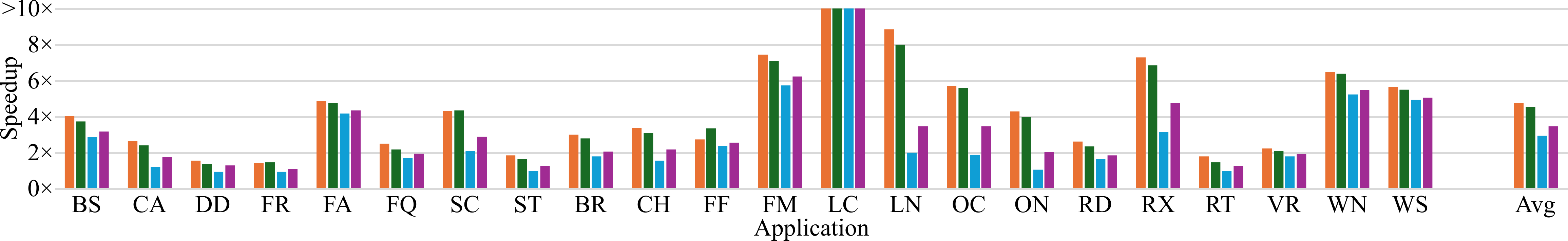}
			\caption{Quad-core configuration}
			\label{sfig:perf_protocol-4}
			\vspace{0.1475cm}
		\end{subfigure}
		\begin{subfigure}[b]{\textwidth}
			\centering
			\includegraphics[width=\textwidth]{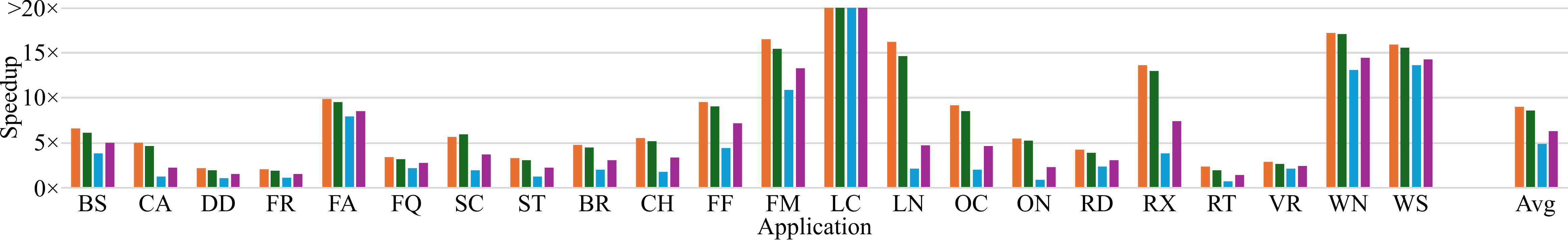}
			\caption{Octa-core configuration}
			\label{sfig:perf_protocol-8}
			\vspace{0.1475cm}
		\end{subfigure}
		\begin{subfigure}[b]{\textwidth}
			\centering
			\includegraphics[width=\textwidth]{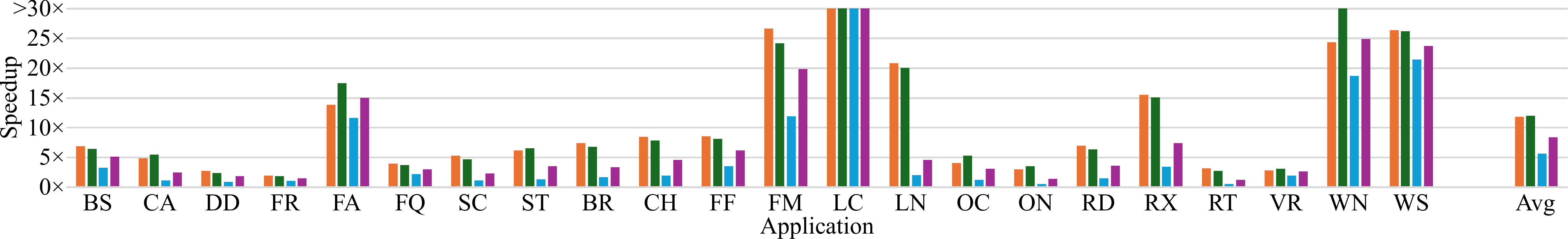}
			\caption{Sixteen-core configuration}
			\label{sfig:perf_protocol-16}
		\end{subfigure}
	\end{minipage}
	\caption{Speedup. i.e., normalized execution time in the simulated system,
		with different applications, numbers of cores, and cache-coherent memory subsystems.
		The system is configured with two, four, eight, or sixteen cores and
		with either gem5 Ruby MI, MESI, or MOESI cache coherence or
		the RTL MSI single- and two-level cache-coherent memory subsystems.
		Time is normalized with respect to the execution with gem5 Ruby's MI coherence protocol.}
	\label{fig:perf_protocol}
\end{figure*}

\subsection{Experimental Results}
\label{ssec:exp_results}
We evaluate the quality of both the RTL/system-level co-simulation
and the use-case MSI cache-coherent memory subsystems designed through
the adoption of the proposed system-level validation flow.
The cache-coherent memory subsystem designs are evaluated according to
the execution time of the considered workload applications, while
the effectiveness of the RTL/system-level co-simulation is measured as
the simulation time of the same executions of applications on the target system.

\begin{figure*}
	\begin{minipage}[b]{\textwidth}
		\centering
		\begin{subfigure}[b]{\columnwidth}
			\centering
			\includegraphics[width=\textwidth]{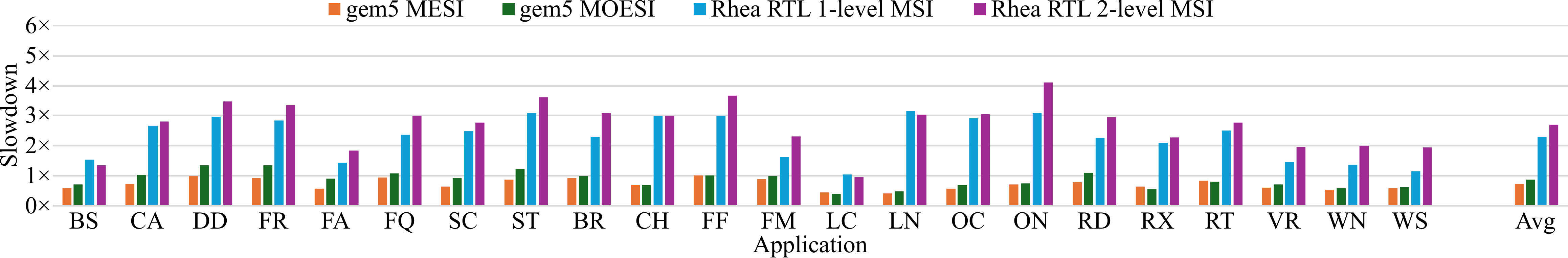}
			\caption{Dual-core configuration}
			\label{sfig:perf_flow-2}
			\vspace{0.1475cm}
		\end{subfigure}
		\begin{subfigure}[b]{\columnwidth}
			\centering
			\includegraphics[width=\textwidth]{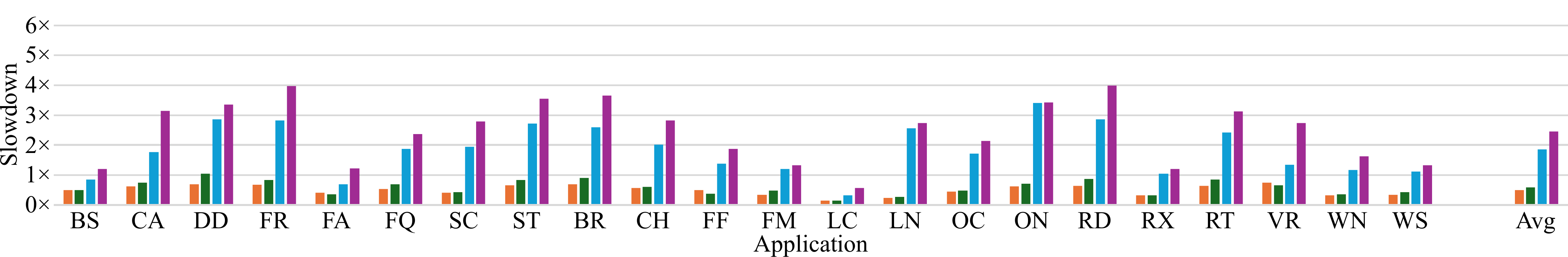}
			\caption{Quad-core configuration}
			\label{sfig:perf_flow-4}
			\vspace{0.1475cm}
		\end{subfigure}
		\begin{subfigure}[b]{\textwidth}
			\centering
			\includegraphics[width=\textwidth]{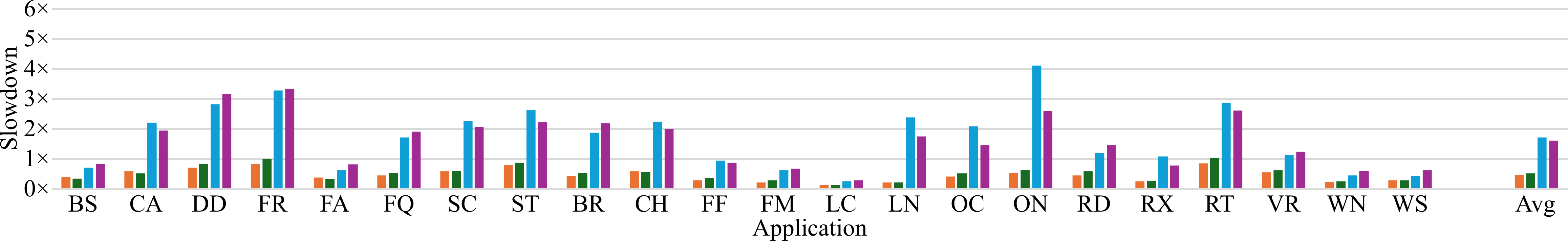}
			\caption{Octa-core configuration}
			\label{sfig:perf_flow-8}
			\vspace{0.1475cm}
		\end{subfigure}
		\begin{subfigure}[b]{\textwidth}
			\centering
			\includegraphics[width=\textwidth]{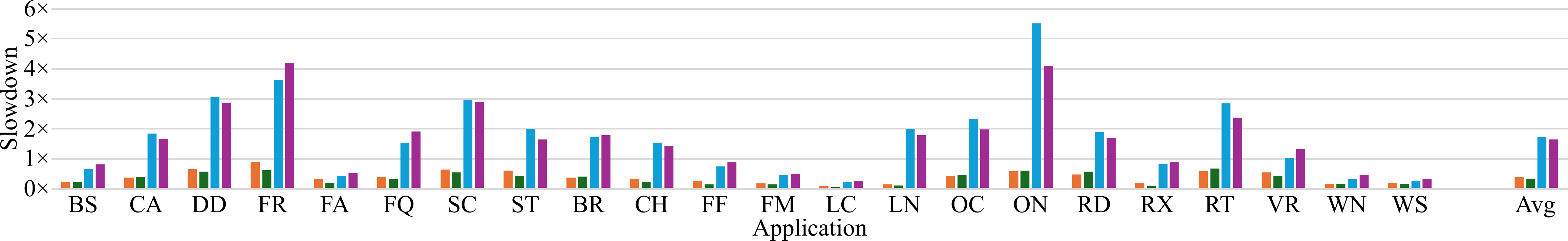}
			\caption{Sixteen-core configuration}
			\label{sfig:perf_flow-16}
		\end{subfigure}
	\end{minipage}
	\caption{Slowdown. i.e., normalized simulation time, of traditional gem5 simulations and of the proposed co-simulation flow
		with different applications, numbers of cores, and cache coherence protocols.
		Traditional gem5 simulations make use of Ruby's MI, MESI, and MOESI cache coherence models,
		while the proposed co-simulation flow simulates a system that includes the obtained
		RTL memory subsystems with MSI cache coherence.
		Time is normalized with respect to gem5 Ruby MI simulations.
	}
	\label{fig:perf_flow}
\end{figure*}

\subsubsection{Cache-Coherent Memory Subsystem Performance}
\label{sssec:exp_results-exec}
We compare the execution time of applications running on
the simulated system depending on the cache-coherent memory subsystem,
which may either implement Ruby's MI, MESI, or MOESI models or
be the MSI-coherent single- and two-level RTL designs obtained by employing the proposed framework.

Simulations are performed on various applications and with different numbers of cores,
and Figure~\ref{fig:perf_protocol} depicts their execution time in the simulated system,
normalized with respect to the corresponding execution time in Ruby's MI scenario.
We refer to such normalized execution time as speedup.
Figures~\ref{sfig:perf_protocol-2}, \ref{sfig:perf_protocol-4}, \ref{sfig:perf_protocol-8}, and \ref{sfig:perf_protocol-16}
respectively chart the speedup when the system features two, four, eight, and sixteen cores.

The experiments demonstrate the ability to evaluate the performance of
the RTL cache-coherent memory subsystem by executing a large number of applications
from the PARSEC and Splash-3 benchmark suites, which are
the de facto standard reference in the field of cycle-accurate simulators.
Moreover, we prove the scalability of the RTL cache-coherent memory subsystem
as well as of the co-simulation flow, both of which are evaluated when
the system features between two and sixteen CPU cores.

The experimental results in Figure~\ref{fig:perf_protocol} show that
both the single- and two-level RTL cache-coherent memory subsystems,
which implement an MSI cache coherence protocol,
perform in line with the expectations, delivering
execution times with intermediate values between those obtained from
the gem5 Ruby's simpler MI and more complex MESI and MOESI with all of
the considered numbers of cores in the overall system, i.e., two, four, eight, and sixteen.
Compared to the simplest MI protocol, the gem5 MESI and MOESI ones and our single-level RTL design
deliver an average speedup that ranges from 2.1\texttimes, 2.0\texttimes, and 1.4\texttimes\, in the dual-core scenario
up to 11.8\texttimes, 12.0\texttimes, and 5.7\texttimes\, in the sixteen-core one, respectively,
as summarized in the rightmost bars of Figure~\ref{fig:perf_protocol}.

The two-level RTL design provides a further performance improvement over the single-level RTL one,
with a reduction in the average execution time of 11\%, 19\%, 33\%, and 43\% when
the system features a dual-, quad-, octa-, and sixteen-core CPU, respectively.
As expected, the benefit of adopting a two-level cache hierarchy is more pronounced
with higher core counts, due to the greater pressure on shared memory resources.

\subsubsection{Co-Simulation Flow Performance}
\label{sssec:exp_results-sim}
We evaluate then the time taken to run simulations of various applications
through the proposed flow and targeting the use-case RTL MSI cache-coherent memory subsystems,
comparing it to traditional gem5 simulations of the same overall system but with
cache-coherent memory subsystems that implement gem5 Ruby's MI, MESI, and MOESI models.

The experiments are performed on each of the 22 considered applications from the PARSEC and Splash-3 suites.
The execution time of each simulation, normalized to the corresponding plain gem5 ones with MI cache coherence
and denoted as slowdown,
is depicted in Figure~\ref{fig:perf_flow}, respectively when the system features two, four, eight, and sixteen CPU cores
in Figures~\ref{sfig:perf_flow-2}, \ref{sfig:perf_flow-4}, \ref{sfig:perf_flow-8}, and \ref{sfig:perf_flow-16}.

Our co-simulation flow is shown to scale effectively as the number of core increases,
with the slowdown compared to gem5 MI simulations that decreases, for the single-level MSI RTL designs,
from 2.3\texttimes\, in the dual-core scenario down to
1.9\texttimes, 1.7\texttimes, and 1.7\texttimes\, in the quad-, octa-, and sixteen-core ones.
Simulations of the two-level MSI RTL designs perform similarly, with a slowdown ranging between 
2.7\texttimes\, and 1.6\texttimes\, and notably lower than the one-level ones for the octa- and sixteen-core scenarios.
Moreover, hybrid gem5-Verilator simulations of single- and two-level MSI RTL designs are both, for any number of CPU cores,
less than 4.9\texttimes\, slower on average
than plain gem5 MESI and MOESI simulations, which are favored by the shorter execution time of the simulated applications.

\subsection{Comparison with State-of-the-Art Approaches}
\label{ssec:exp_comparison}
Remarkably, the simulation time overhead incurred by the proposed gem5-Verilator co-simulation flow is
orders of magnitude smaller than that of traditional RTL simulators,
e.g., Cadence Xcelium.
The latter, despite their fine-grained accuracy, entail significantly longer simulation times
and do not readily support the rapid setup of a complete system capable of booting an OS
and executing real-world applications.

Stimulating the RTL design with the execution of such applications,
particularly during the early stages of the design process,
is only feasible through the use of a system-level simulator such as gem5.
Importantly, the moderate overhead of the hybrid co-simulation is offset by the drastically increased fidelity it provides.
This enhanced accuracy stems from conducting the evaluation on the actual RTL design of the cache-coherent memory subsystem,
rather than relying on custom gem5 models or generic gem5 cache coherence protocol implementations.

Finally, we note that the hybrid gem5-Verilator co-simulation flow integrated in Rhea performs on par, from a simulation time overhead standpoint, with gem5+RTL~\cite{Lopez-Paradis_2021ICPP}.
Although also integrating the gem5 and Verilator tools, gem5+RTL primarily targets however hardware accelerators, e.g., Nvidia's NVDLA deep-learning accelerator, or much simpler components and lacks instead dedicated support for modeling or validating complex RTL cache-coherent memory subsystems.

In contrast, the Rhea framework introduced in this manuscript provides a complete design and validation solution tailored for cache-coherent memory subsystems, also including RTL generation of configurable designs and the SystemVerilog port of gem5’s Ruby tester for additional standalone protocol-level stress testing of the RTL design.

\section{Conclusions}
\label{sec:conclusions}
This paper introduced Rhea, a framework that enables the fast design and system-level validation of
RTL cache-coherent memory subsystems, notably combining gem5's system-level and Verilator's RTL
simulation to verify their functional correctness and evaluate their performance.

As use cases, we applied the Rhea framework to design
RTL cache-coherent memory subsystems implementing the MSI cache coherence protocol
with one and two levels of caches and supporting up to sixteen CPU cores.

An extensive experimental campaign has demonstrated the quality of
the RTL cache-coherent memory subsystems that can be designed by
adopting the framework and has evaluated the impact given by
integrating Verilator with gem5 to validate the RTL design while stressing it in
a system that can execute real-world applications and run an OS.

The designed MSI cache-coherent memory subsystems were evaluated in dual-, quad-, octa-, and sixteen-core CPU configurations while executing 22 different applications from state-of-the-art benchmark suites, and were shown to deliver an intermediate performance compared to the gem5 Ruby's simpler MI model and the more complex MESI and MOESI.

The proposed RTL/system-level co-simulation flow was shown instead to have a slowdown in terms of simulation time compared to gem5 Ruby MI simulations of only up to 2.7\texttimes, while simulating not a custom gem5 model of the latter or a generic gem5 coherence protocol, but the actual RTL design of the cache-coherent memory subsystem under verification.
Remarkably, the Rhea framework scales effectively as the core count increases, with the slowdown compared to MI simulations decreasing down to 1.6\texttimes\, in octa- and sixteen-core scenarios.

\bibliographystyle{IEEEtran}
\bibliography{2025_arXivRhea}

\end{document}